\documentclass[twocolumn]{aastex701}

\usepackage{graphicx}	
\usepackage{amsmath}

\begin{document}

\title[SMMAN studies of AGN variability. II. Radio and optical quasi-periodic oscillations and changing variability states in the relativistic jet of the blazar PKS 0754+100  ]{SMMAN studies of AGN variability. II. Radio and optical quasi-periodic oscillations and changing variability states in the relativistic jet of the blazar PKS 0754+100}


\author[0000-0002-3960-5870]{Ashutosh Tripathi}
\affiliation{State Key Laboratory of Radio Astronomy and Technology, Xinjiang Astronomical Observatory, CAS, 150 Science 1-Street, Urumqi 830011, China}
\email[show]{ashutoshtripathi@xao.ac.cn}  

\author[0000-0001-8221-9601]{Xin Wang}
\affiliation{State Key Laboratory of Radio Astronomy and Technology, Xinjiang Astronomical Observatory, CAS, 150 Science 1-Street, Urumqi 830011, China}
\email{wangxin2019@xao.ac.cn}

\author[0009-0003-6680-1628]{Xi Yan}
\affiliation{State Key Laboratory of Radio Astronomy and Technology, Xinjiang Astronomical Observatory, CAS, 150 Science 1-Street, Urumqi 830011, China}
\email{yanxi@xao.ac.cn}

\author[0000-0002-1029-3746]{Paul J. Wiita}
\affiliation{Department of Physics, The College of New Jersey, 2000 Pennington Rd., Ewing, New Jersey 08628-0718, USA}
\email{wiitap@tcnj.edu}

\author[0000-0003-0721-5509]{Lang Cui}
\affiliation{State Key Laboratory of Radio Astronomy and Technology, Xinjiang Astronomical Observatory, CAS, 150 Science 1-Street, Urumqi 830011, China}
\affiliation{Xinjiang Key Laboratory of Radio Astrophysics, 150 Science 1-Street, Urumqi, Xinjiang, 830011, China}
\email[show]{cuilang@xao.ac.cn}

\begin{abstract}
We report the detection of $\sim$280-day quasi-periodic oscillations (QPOs) in the SMMAN observations of PKS 0754+100 at 4.8 and 23.6 GHz, which may also be present in VLBI observations made at 15 GHz. The onset of a transition of the amplitude of variability in the radio flux density, observed more clearly in the 15 GHz and 23.6 GHz observations, coincides with the detection of a likely QPO of $\sim$1.4 days in optical observations made by the Transiting Exoplanet Survey Satellite. This transition in radio variability also occurs close in time to the disappearance of the secondary knot in the VLBI radio map during the lower variability state. An X-ray flare was observed 100 days before this transition, suggesting that the mechanisms causing these quasi-periodicities are related. After the transition to a higher variability state, both the 15 and 23.6 GHz light curves exhibit indications of QPOs at $\sim$400 days. For the first time, a QPO detected in the radio band from a BL Lac has been seen to appear to shift to a QPO of lower frequency, with that transition occurring near the appearance of a QPO in optical emission at a much shorter period. This multi-waveband behavior could be related to the internal helical structure of the jet, where an instability or kink develops due to magnetic reconnection and propagates outward. 
\end{abstract}

\section{Introduction} 
Active galactic Nuclei (AGNs) are extremely luminous objects, with powers sometimes exceeding $10^{47}$ erg s$^{-1}$, and are understood to contain supermassive black holes with masses in the range $10^6$--$10^{10}$ $M_\odot$ \citep{2017A&ARv..25....2P} and are usually classified using their radio and optical properties \citep{1989AJ.....98.1195K}. Around one-tenth of them eject relativistic jets and they produce powerful radio emission from the synchrotron process, with a strength depending on the magnetic field and density of ultra-relativistic electrons in the jets \citep{1995PASP..107..803U}. Blazars are brighter AGNs with relativistic jets pointing close to our line of sight, which causes the Doppler-boosted emission from jets to dominate the whole emission of the galaxy and also produces amplified variability throughout the electromagnetic spectrum \citep{1997ARA&A..35..445U, 2025A&ARv..33....8R}. Blazars are subdivided into two classes: BL Lacertae objects (BL Lacs) and flat-spectrum radio quasars (FSRQs). In addition to jet emission, FSRQs also show strong emission lines \citep{1978PhyS...17..265B, 1997A&A...327...61G}, whereas BL Lac objects show no or weak emission lines with almost all emission coming from the relativistic jet \citep{1991ApJ...374...72S, 1996MNRAS.281..425M}.

Variability is characteristic of the light curves of blazars throughout the electromagnetic spectrum and has proven to be effective in understanding their structure and radiative processes \citep{1979ApJ...232...34B, 2025A&ARv..33....8R}. These variations are usually aperiodic and are attributed to stochastic variations inside jets, especially in the optical and radio wavebands \citep{1985ApJ...298..114M, 2008Natur.452..966M}. Quasi-periodic oscillations (QPOs) correspond to essentially periodic and possibly deterministic variations, which have rarely been clearly observed in blazars \citep[e.g.][]{2008Natur.455..369G, 2022Natur.609..265J}. When radio QPOs are detected, their timescales are found to be of the order of years and are believed to be related to the presence of a binary supermassive black hole system in some cases, and more generally to the approaching jet, either because it is precessing or is subject to internal processes such as a kink instability \citep[e.g.][]{2021MNRAS.501.5997T, 2022ApJ...926L..35O}. 

Although there are many claims of QPOs in some multi-waveband observations of blazars, there are intrinsic difficulties which make many of these detections questionable. Unlike stellar-mass black hole binaries, QPOs in AGNs usually show relatively fewer cycles due to the much longer QPO periods associated with them \citep{2024ApJ...977..166T, 2026ApJ...996L..39R}. For instance, claimed radio QPOs usually have an order of a few years \citep{2021MNRAS.501.5997T, 2022ApJ...926L..35O}. In order to observe completely convincing sets of 15--20 cycles, the temporal baseline of observation should be at least 60 years, which is far longer than any existing monitoring program \citep{2026ApJS..283...69C}. Another issue is the presence of complex stochastic variations that, if not properly accounted for, can cause distortion in the QPO signal and may produce false ones \citep[e.g.][]{2016MNRAS.461.3145V}.

PKS 0754+100 is a BL Lac situated at the redshift of $z=0.266$ \citep{2003A&A...412..651C} and its extreme variability has been extensively studied in X-rays \citep{1990ApJ...356..432G, 2013RAA....13..259F}, optical \citep{2004A&A...418..813H, 2000ApJS..127...11G, 2016A&A...591A..21M}, radio \citep{2007A&A...462..547F, 2009AJ....137.5022N}, and $\gamma$-rays \citep{2011MNRAS.414.2674G, 2025MNRAS.540.1449A}. This source has also been part of multiple observations with the Very Long Baseline Array (VLBA) \citep{2007A&A...466...63W, 2016AJ....152...12L} and a program triggered by neutrino detections which produces real-time alerts for follow-up $\gamma$-ray observations \citep{2016JInst..1111009I}.

Until now, no QPOs have been observed in any wavebands for this source except for one reported by \citet{2002A&A...381....1F} in  optical B-band observations spanning 82 years with a possible period of around 18 years.  However, this has too few cycles to be considered a QPO in that 3 or 4  ``cycles'' can  arise from red-noise \citep{1978ComAp...7..103P, 2016MNRAS.461.3145V}. They also noted a possible 3-year QPO in a segment of these observations. In this work, we report for the first time good evidence for multi-frequency flux quasi-periodicities at radio and optical wavelengths. 
 
\begin{figure*}
    
   \includegraphics[width=1.0\linewidth]{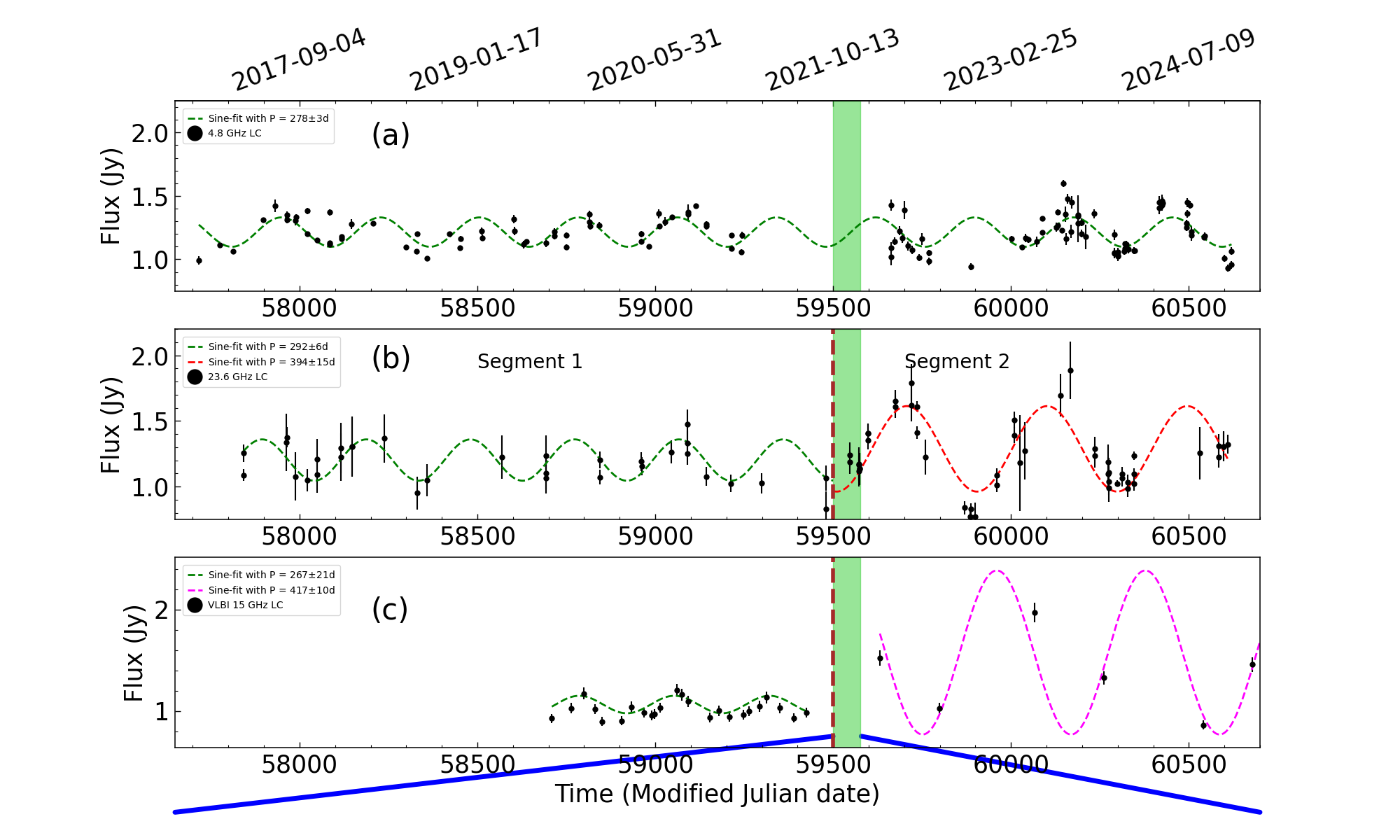}
   \includegraphics[width=1.0\linewidth]{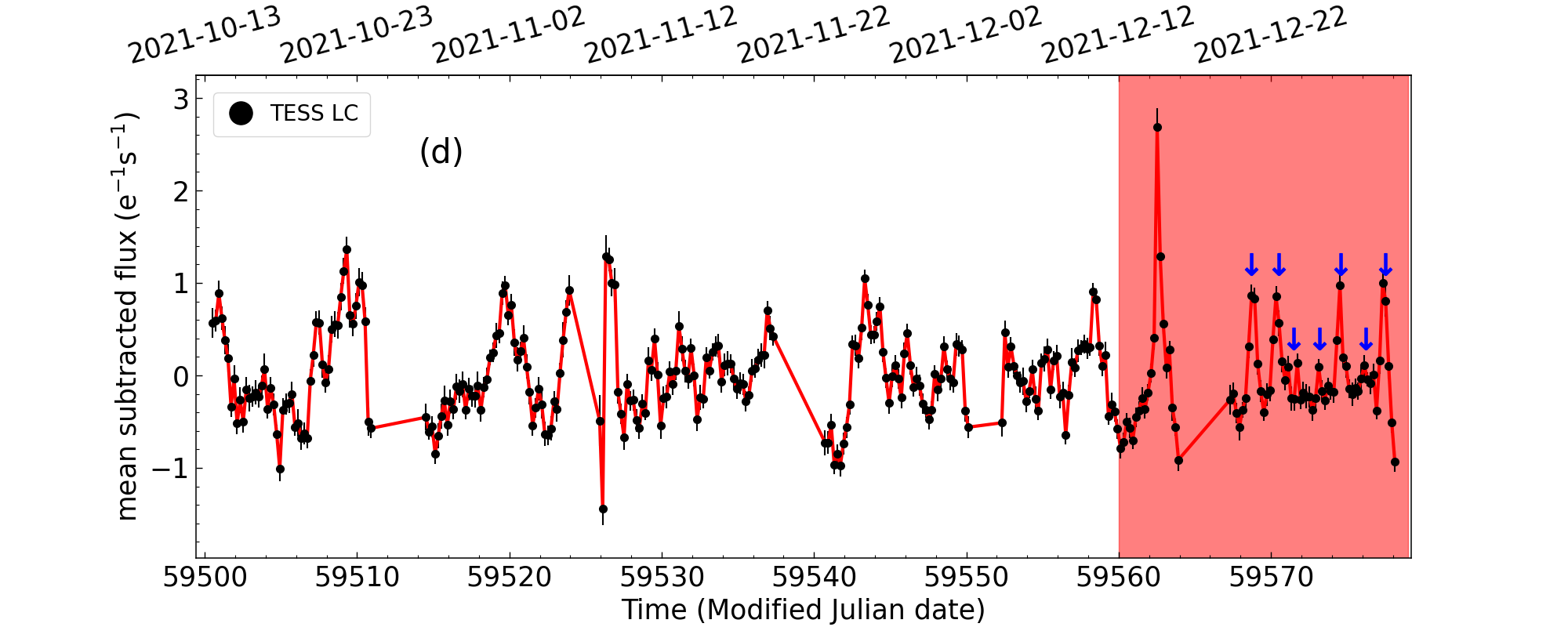}
    \caption{SMMAN and TESS observations of PKS 0754+100: The black dots denote the flux density measurements with 1$\sigma$ error bars. Panels (a) and (b) plot the light curves at 4.8 GHz and 23.6 GHz, respectively. The green dashed line plots the fit of the observations with sine functions of periods of 278$\pm$3 days and 292$\pm$6 days at 4.8 GHz and 23.6 GHz, respectively; the periods and amplitude of variations of sine curves plotted in this figure are obtained using GLSP. For 23.6 GHz, the period after MJD 59500 is fit with the period of 394$\pm$15 days and is shown with a red dashed line. Panel (c) plots the VLBI flux measurements during the period of SMMAN observations, and the green curve denotes sinusoidal oscillations with a period of 267$\pm$21 days. The observations after MJD 59500 show a possible periodic modulation of 417$\pm$10 days in the dashed curve in magenta. The dashed brown line marks the onset of the higher variability state seen in the 15 and 23.6 GHz light curves, which we use to separate the data into segments 1 and 2 with lower and higher variability, respectively. The shaded region in panels (a), (b), and (c) marks the period MJD 59500--59577 during which simultaneous TESS observations were made. Panel (d) shows the 5-hour binned TESS light curve from sectors 44, 45, and 46. The shaded region in red shows the series of peaks, marked by arrows, which indicate a possible QPO at $1.4 \pm 0.1$ day. }
    \label{fig-lc}
\end{figure*}

\begin{figure*}
\centering
   \includegraphics[width=0.7\linewidth]
   {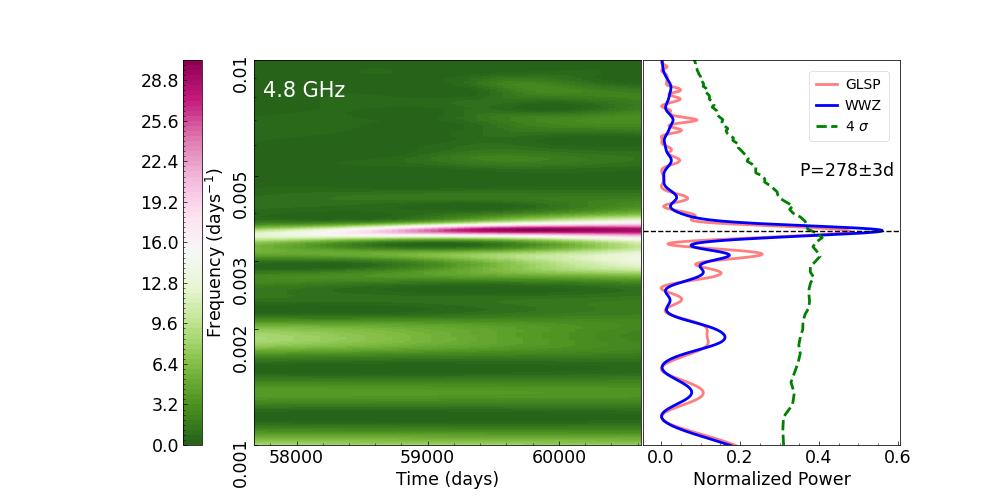}\hspace{-1cm}\includegraphics[width=0.7\linewidth]{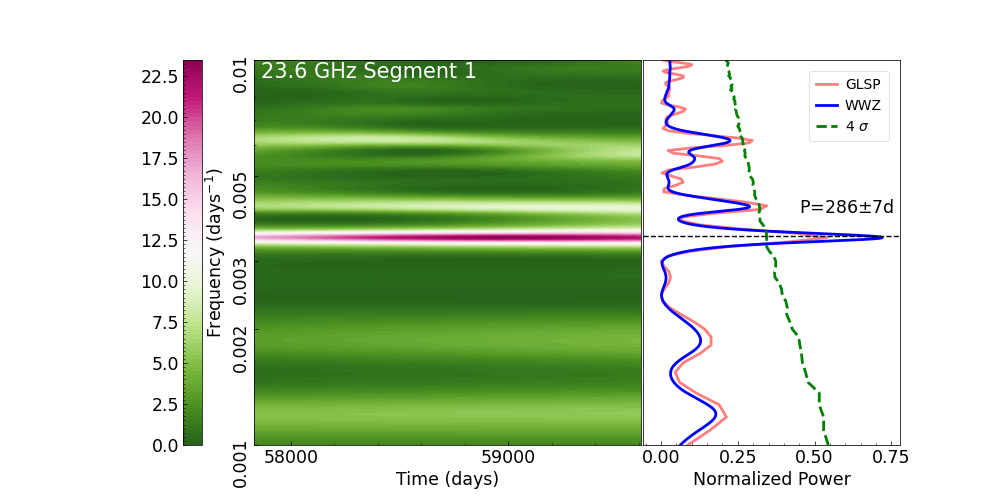}
    \includegraphics[width=0.7\linewidth]{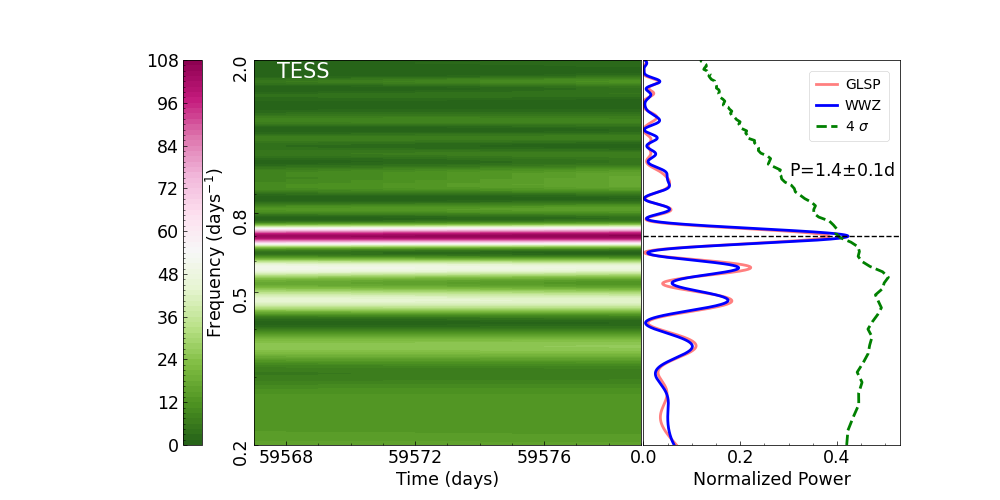}
        \caption{Analysis of radio and optical observations: The left plot of each panel shows the weighted Wavelet Z-transform (WWZ) diagram with the power decreasing as colour advances from magenta to green. The right plot shows the generalized Lomb-Scargle periodogram (GLSP) and time-averaged WWZ plotted in red and blue, respectively. The corresponding 4$\sigma$ local significance (99.99\%) levels are denoted by the dashed green curves. The QPO periods quoted come from the time-averaged WWZ. The top panel shows the result for 4.8 GHz between MJD 57700 and 60620.  The GLSP shows a prominent peak around 278 days with 4.51$\sigma$ local significance and a strong concentration of WWZ power is observed at the frequency of 0.0036 d$^{-1}$, also corresponding to the period of 278 days. The middle panel shows the results for 23.6 GHz for segment 1 (MJD 57700--59500), where a peak in GLSP is observed at 283 $\pm$ 5 days having local significance of 4.89$\sigma$ and WWZ shows most of the power concentrated around 286 days,  and it strengthens as the observation progresses towards MJD 59500. The bottom panel shows the WWZ and GLSP analyses for the TESS data between MJD 59568 and 59578. The WWZ plot shows a very strong peak throughout the observation, at a period of 1.4 days. The GLSP and time-averaged WWZ plotted in the right panel also show that peak, with local significance of 4.72$\sigma$. }
    \label{fig-res}
\end{figure*}

\section{QPO analysis}
This source is part of the quasi-Simultaneous Multiwavelength Monitoring of gamma-ray-loud AGNs with the Nanshan 26-m radio telescope (SMMAN program) operating at frequencies of 4.8 and 23.6 GHz \citep{2026ApJS..283...69C, wang2026smmanstudiesagnvariability}. In the SMMAN sample, there are 131 sources for which measurements at both 4.8 and 23.6 GHz are available.  In examining these radio observations we found only 11 sources to be noticeably variable. We were not particularly looking for QPOs and focused on PKS 0754+100 because it showed particularly interesting variability. As discussed below, we also found interesting variability in the optical behavior as well as an X-ray flare that might be related to a single physical mechanism.

The decade-long SMMAN light curves of PKS 0754+100 at 4.8 and 23.6 GHz, together with the 15 GHz light curve measured with the VLBA \citep{2018ApJS..234...12L} are shown in Figure \ref{fig-lc}. Long-term variability has been detrended from these light curves, as it will affect Fourier and wavelet analysis at lower frequencies \citep{2024ApJ...966..158P}; see Appendix A. The mean flux densities  of the 4.8, 15, and 23.6 GHz light curves are 1.19, 1.08, and 1.20 Jy, respectively. The root mean square (RMS) of the flux at 4.8 GHz is $\sim$ 0.1, which is also the case for the first segment (before MJD 59500) of the 23.6 GHz light curve. After MJD 59500, the RMS of the 23.6 GHz light curve increases by almost a factor of three, to 0.28 as seen in Figure \ref{fig-lc}. Similar behavior is also exhibited in the 15 GHz light curve, where the variation in the flux density measurements clearly increases after MJD 59500 despite the relative paucity of the VLBI observations available for this dataset. During the lower variability state, nearly periodic modulations of similar lengths appear to be present at all three radio frequencies, as is indicated by the sinusoidal components shown in green with a period of 278, 267, and 292 days for 4.8, 15, and 23.6 GHz, respectively. During the latter portion of these observations  with higher variability the flux densities show tentative periodicities of 410 and 417 days at 23.6 and 15 GHz, respectively.  The Transiting Exoplanet Survey Satellite (TESS) also observed this source for three consecutive sectors during October--December 2021 as shown in panel (d) of Fig.~\ref{fig-lc}, where periodic modulations in the optical band also appear to be present during the MJD 59568--59578 following a large flare beginning on MJD 59560.

Figure~\ref{fig-res} shows our analysis of light curves using the generalized Lomb-Scargle periodogram (GLSP) \citep{1976Ap&SS..39..447L, 1982ApJ...263..835S, 2009A&A...496..577Z} and the weighted Wavelet Z-transform (WWZ) \citep{1996AJ....112.1709F, npg-12-345-2005} techniques. In the weighted wavelet Z-transform (WWZ), wavelet power is a function of both frequency and time, which means that it will describe the behavior of the peak not only in the frequency domain as done by GLSP, but also illustrates any temporal evolution during the observation. Marginalizing the WWZ over time gives its value as a function only of frequency, similar to the periodogram, and is known as time-averaged WWZ. The errors quoted on the QPO periods  are related to the frequency resolution employed in computing the PSD.

The GLSP for 4.8 GHz is plotted in the right plot of the upper panel, indicating a period of 278 $\pm$ 3 days with local significance of 4.51$\sigma$. When the number of frequencies examined and the number of sources checked for possible periodicity are included to take into account the ``look elsewhere effect'', the global significance becomes 2.98$\sigma$.  See Appendix D for details on significance calculations. The temporal baseline of 2900 days yields $\sim$10 oscillation cycles, which is convincing enough to be considered as a QPO \citep[e.g.][]{2016MNRAS.461.3145V}. In the left plot, the WWZ power is most concentrated around the frequency of 0.0036 d$^{-1}$, consistent with what was found with the GLSP analysis. The time-averaged WWZ also shows a strong peak at 278 $\pm$ 2 days. 

The middle panel shows the results for the first segment of the 23.6 GHz light curve. Here, the GLSP shows a peak at 292 $\pm$ 6 days with a local significance of 4.89$\sigma$, producing 6 oscillations over a temporal baseline of 1725 days. The local significance becomes 3.5$\sigma$ after including the look elsewhere effects. The WWZ analysis shows a strong peak at 286 $\pm$ 7 days, which is somewhat weaker at the beginning of the observation but gains strength towards the end. The slightly higher QPO period from the GLSP approach could be attributed to the larger flux density errors and is consistent within the statistical errors with the WWZ peak at 286 $\pm$ 7 days. The latter is within the errors of the 4.8 GHz peak at 278 $\pm$ 3 days,  which is the same from both the GLSP and WWZ analyzes.  
The WWZ plot for the entire light curve at 23.6 GHz (Fig.~\ref{fig:k}) shows a stronger signal around 494 days while the signal around 290 days is weakened, implying that the transition in state at 23.6 GHz radio frequency when the variability amplitude increases introduces a solid signal at the longer period.

\begin{figure*}
    \centering
    \includegraphics[width=1.0\linewidth]{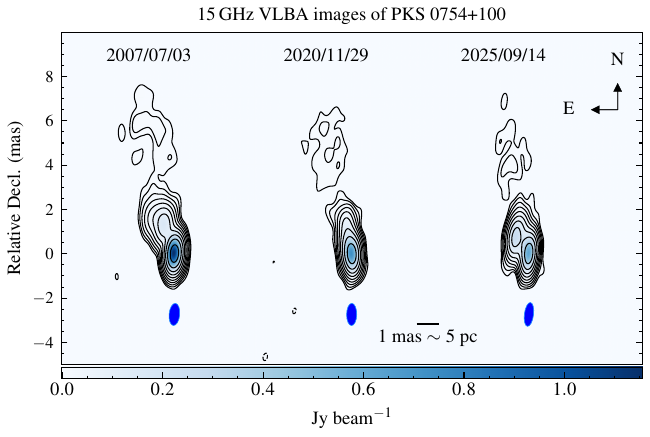}
    \caption{15 GHz Very Long Baseline Array (VLBA) radio images for selected epochs between 2007--2025 for PKS 0754+100: The left image was obtained during an earlier high variability state and both a
    prominent core and a knot component are seen. Similar behavior is also observed for the radio image taken on 2025/09/14 during the high-variability state shown in Fig.~\ref{fig-lc}. However, during the low-variability state (middle image) no distinctive knot components are present and the central component appears as one extended core. These VLBI radio maps are obtained from the calibrated data of this source available on the MOJAVE \citep{2018ApJS..234...12L} website; see Appendix C.}
    \label{fig-map}
\end{figure*}

The 15 GHz VLBA light curve \citep{2018ApJS..234...12L} possesses a possible periodic modulation with a period of $\sim$267 $\pm$ 21 days during the interval MJD 58709--59425, as shown in Fig.~\ref{fig-lc}. As this possible period is less than 3 oscillations for the temporal baseline of around 720 days, it is not enough for the signal to be considered as a QPO. However, the presence of QPO signals of similar periods in the other radio frequencies observed at Nanshan during the same overall interval makes it worthwhile to include these data in our discussion, even without VLBA observations made at other epochs overlapping with the SMMAN observations. Although QPO signals at 4.8 and 23.6 GHz have global significances of around 3$\sigma$, the robustness of these QPO signals is supported by independent VLBI flux measurements made at 15 GHz. The modest global significances could arise from the presence of some long-term trends or noise in the radio flux measurements.

The bottom panel of Figure \ref{fig-res} shows the analysis of the TESS observations of PKS 0754+100 for the interval MJD 59568--59578. Although strong optical variability is present in the entire length of Sector 46 (MJD 59550--59578), we note that the optical flare that occurred in the middle of Sector 46 (around MJD 59562) may have triggered subsequent apparent periodic patterns in the observation, starting around MJD 59598. The WWZ diagram shows most of the power around  0.7 d$^{-1}$, which persists throughout the observations, corresponding to a period of 1.4 $\pm$ 0.1 days. The GLSP and time-averaged WWZ also show prominent peaks around 1.4 d with a local (global) significance of 3.85~$\sigma$ (2.87~$\sigma$). This 1.4 d period yields 7 oscillations for the baseline of 10 days, which is normally considered to be enough to be considered as a QPO. The peaks of these oscillations are marked in the  TESS light curves shown in Fig.~\ref{fig-lc}.

\section{Discussion}

Our SMMAN observations of PKS 0754+100 made at 4.8 and 23.6 GHz during the low-variability state indicate the presence of a $\sim$280 d QPO, which is supported by the VLBI flux density measurements. The transition from a low to a high radio variability state is accompanied by a $\sim$1.4 d QPO seen in the TESS observations which follow an X-ray flare observed by Swift.

\subsection{Transition in variability state}

To understand the origin of the physical process responsible for the change in the variability state of PKS 0754+100 in the radio bands, we obtained VLBI radio images during different epochs, as shown in Fig.~\ref{fig-map} (see Appendix C on VLBA data reduction). During the initial high state, the radio morphology is characterized by a bright compact core accompanied by well-defined, prominent knot components, which are observed to move outward from the central core \citep[see the movies from the MOJAVE program;][] {2018ApJS..234...12L}. In contrast, the low variability state (second image) is dominated by a comparatively smoother and more compact radio structure. 
In the third map, when the variability becomes higher again, the central core appears to have a second component, which, however, is not as distinctive as in the initial epoch shown here. So, the emission structure changes during the low variability state, and the single-central core suggests that more of the emission originates closer to the central black hole. The radio morphology changes around MJD 59500 and the fact that the flux becomes more highly variable afterward indicate a possible connection between these observations. Furthermore, the position angle of the outward-moving knot components appears to vary significantly over time. However, due to the sparse temporal sampling of the VLBA available observations and the strong dependence of brightness on viewing angle, we are unable to determine whether this apparent variation is associated with jet precession, jet bending, or another physical mechanism.

The 15 GHz VLBA light curve \citep{2018ApJS..234...12L} and the 15--50 keV X-ray light curve from the Swift satellite \citep{2013ApJS..209...14K} have been obtained for this source; these data cover the epochs after July 1995 and February 2005, respectively (see Appendix C and Fig.~\ref{fig-swift}). Before the large gap in VLBA observations between January 2012 and August 2019, the 15 GHz light curve exhibited a very high variability state. After this large gap, it was in a lower variability state, which seems to transition to a higher one again after October 2021 (MJD 59500), strikingly similar to the 23.6 GHz fluxes. The Swift X-ray light curve is nearly continuous and shows relatively steady fluxes between 2005 and 2021, but starts to show some significant variations starting at MJD 59400, which is around the onset of the high-variability state in the 15 and 23.6 GHz observations. So, the transition of the VLBI radio variability from low to high states is also apparently reflected in the X-ray Swift light curves which exhibited stronger flares; this may imply that the change in the radio flux originates very close to the black hole. The Swift data are, however, too noisy to allow for sensible GLSP or WWZ analyzes. We note that the periodic modulation observed in the TESS observations during MJD 59500--59578 is observed shortly after the X-ray flare starting at MJD 59410 and also around the time of the transition to a high-variability state in the radio observations after MJD 59500.

\subsection{Theoretical models}

Some radio QPOs, with periods of several years, can be explained through the presence of binary supermassive black hole AGN systems \citep{1996ApJ...460..207L, 2007A&A...462..547F, 2018MNRAS.478..359K, 2024ApJ...968L..17V}. However, this scenario is most plausible when much of the enhanced QPO emission comes from the accretion disk, which is very unlikely for BL Lac objects such as PKS 0754+100, where most of the emissions almost certainly come from the Doppler boosted relativistic jet. Also, as this binary supermassive black hole scenario produces modulations at timescales that differ significantly from the radio QPOs found in this work, we do not expect the QPOs found at the radio frequencies for PKS 0754+100 to arise this way.

\begin{figure*}
    \centering
    \includegraphics[width=1.0\linewidth]{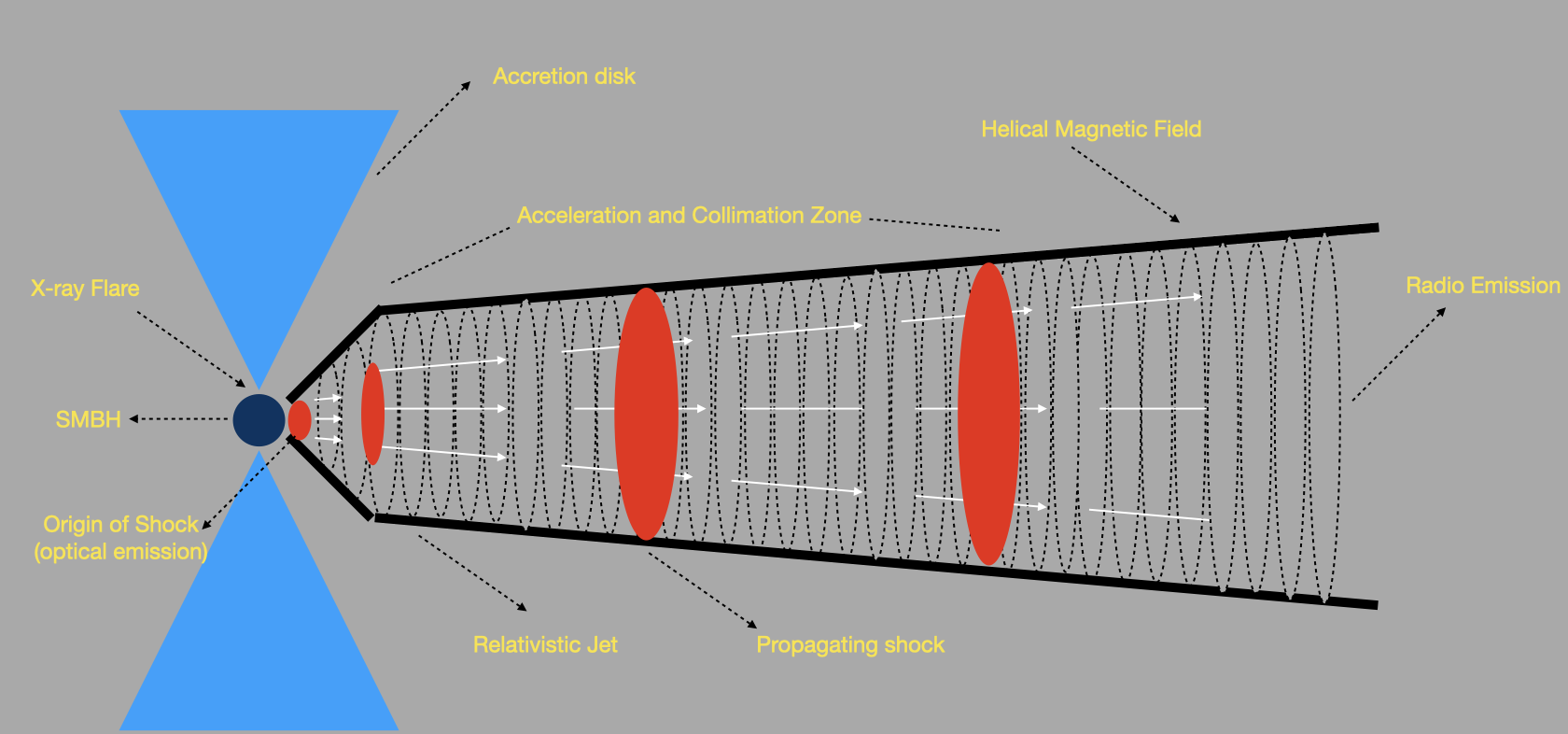}
    \caption{Schematic illustration of the origin and development of QPOs. The X-ray emission originates in the innermost region of AGN and then propagates to the base of the relativistic jet, which produces the observed optical modulations. This optical modulation grows outwards towards the jets and affects the radio observations, which modifies the existing periodic modulation and also shows periodic modulation of a longer period. }
    \label{fig-sketch}
\end{figure*}

Precession of the jets arising from the inner portion of the accretion disk around the supermassive black hole through frame-dragging (Lense-Thirring precession) can produce a quasi-periodicity of the order of weeks to months in AGN \citep{1998ApJ...492L..59S, 2018ApJS..234...12L}. This model has been used to explain the observed 600 d period at 22.0 and 37 GHz observations for 3C 454.3 \citep{2024ApJ...977..166T}. The QPOs we have found here for PKS 0754+100, with periods $\sim$280 days in all three radio bands, could thus be produced via jet precession. Another possible model is related to the internal helical structure of the jet \citep{2004ApJ...615L...5R, 2015ApJ...805...91M}. Instabilities in the current-driven plasma produce a propagating shock which interacts with the toroidal magnetic field in the jet. This interaction distorts the magnetic field around the jets, producing  modulations in the observed flux \citep[e.g.][]{1992A&A...255...59C, 2004ApJ...601..414L}. Recall that the low variability state of PKS 0754+100 is accompanied by the absence of a secondary moving knot in the VLBI total intensity image at 15 GHz. The similar disappearance of moving knots in 3C 454.3 has been explained in terms of the bending of a jet or the presence of a standing shock \citep{2024A&A...682A.154T}. So, there are two generic scenarios: the first one is related to the geometry of the jet, and the second to instabilities within, or shocks propagating along, the jet. 

After MJD 59500, the radio fluxes at 15.0 and 23.6 GHz start to change from a low to a high variability state, resulting in possibly detected high-amplitude QPOs with periods around 400 days. It therefore appears that the physical process (flaring activity or relativistic beaming) causing this change would affect the quasi-periodic behavior of the higher frequency radio fluxes without abruptly interrupting the ongoing process itself.  High variability is also observed in optical TESS observations, starting MJD 59500 for around 80 days. A QPO of 1.4 days in the TESS light curve starting at MJD 59565 suggests a connection between the physical processes producing strong modulation in both radio and optical wavebands. This optical quasi-periodicity also has two possible scenarios related to geometry and shocks, similar to those for radio observations. Shocks, which propagate along the jet, interact with the toroidal magnetic field of the jet and can produce a kink instability, as seen in magneto-hydrodynamic simulations \citep{2012MNRAS.423.3083M, 2020MNRAS.494.1817D}. The temporal growth of this kink mode from the center of the jet to its boundary could be seen as a periodic flux modulation of the order of a few hours to days \citep{2022Natur.609..265J, 2024MNRAS.527.9132T, 2024MNRAS.528.6608T}. Another possibility related to the geometry of the jet is the presence of sub-structure (``mini-jets") \citep{2012A&A...548A.123B, 2020ApJ...903..134C} which precess along with the main jet and produce a periodicity of the order of a few days \citep{2026MNRAS.545f1920T, 2026MNRAS.545f2211T}. The reported radio and optical quasi-periodicity can involve an elaboration of the classical shock-in-jet model \citep{1985ApJ...298..114M, 2015A&A...580A..94F}, where a shock travels throughout the conical relativistic jet, producing general radio, optical and X-ray variability.

It is worth noting that a strong flare was also observed in the Swift X-ray observations of PKS 0754+100 during MJD 59400--59440, just before the onset of high radio variability. The subsequent onset of the apparent QPO in the optical TESS data suggests the possibility that the optical variability is related to this X-ray flare. So, the periodic modulation observed after MJD 59500 in both optical and radio bands could arise from a single physical process which originates in the inner-most regions of the AGN and causes flux variations in X-rays, as shown in Fig.~\ref{fig-sketch}. These flux variations propagate into the jet, producing a shock, which eventually interacts with its toroidal magnetic fields, and yielding variations that are manifested in flux modulations in optical and radio wavebands. These variations can be viewed as disk-jet coupling, where changes in the vicinity of the accretion disk are manifested in the jet dynamics. The high-variability state, seen in the later portions of the 15 and 23.6 GHz light curves, appears to show higher amplitude quasi-periodicity with a longer period, suggesting that the propagating shock amplifies and modifies the physical process producing the radio QPO in the low-variability state.

Before MJD 59500, the QPO behavior of PKS 0754+100 can be described by either the geometrical precession related model or a scenario where a shock interacts with internal jet structures. After MJD 59500, the multi-wavelength observation suggests that the change in variability state at 15 and 23.6 GHz is caused by a shock, possibly connected to the X-ray flare around MJD 59400. 

For the first time, this work provides clear evidence in a blazar for QPOs in the optical and radio wavebands of the order of 1 and 280 days, respectively. The apparent change in the quasi-periodicity in the 15 and 23.6 GHz light curves after MJD 59500 can not be confirmed at this stage due to the limited number of cycles seen, and we will search for additional modulation cycles in the ongoing SMMAN and other long-term radio monitoring programs.  \\

\begin{acknowledgments}
TESS data presented in this article were obtained from the Mikulski Archive for Space Telescopes (MAST) at the Space Telescope Science Institute. The specific observations analyzed can be accessed via \dataset[DOI: 10.17909/t9g7-1c78]{https://doi.org/10.17909/t9g7-1c78}. This research has made use of data from the MOJAVE database that is maintained by the MOJAVE team \citep{2018ApJS..234...12L}. The Very Long Baseline Array is operated by the National Radio Astronomy Observatory. The National Radio Astronomy Observatory and Green Bank Observatory are facilities of the U.S. National Science Foundation operated under cooperative agreement by Associated Universities, Inc.  

This work is supported by the Tianshan Talent Training Program (Grant No.\ 2023TSYCCX0099) and the National Key R\&D Program of China (Grant Nos.\ 2024YFA1611503 and 2022SKA0120102). A.T. and X.Y. acknowledge support from the Xinjiang Tianchi Talent Program. X.Y. is also supported by the China Postdoctoral Science Foundation under grant Nos.\ 2025M773200 and 2026T190850. This work is also partly supported by the Urumqi Nanshan Astronomy and Deep Space Exploration Observation and Research Station of Xinjiang (XJYWZ2303), the Central Guidance for Local Science and Technology Development Fund (Grant No.\ ZYYD2026JD01), and the Astrometric Reference Frame project (Grant No.\ JZZX-0102). 

\end{acknowledgments}

\appendix
\section{SMMAN data} 
The observations of PKS 0754+100 at 4.8 and 23.6 GHz have been collected by the Nanshan 26 m telescope of the Xinjiang Astronomical Observatory in our ongoing long-term monitoring of radio-loud AGNs (SMMAN program, \citep{2026ApJS..283...69C, wang2026smmanstudiesagnvariability}) which began in November 2016. The receiver used in this monitoring  operates in bands centered around 4800 MHz and 23600 MHz, with bandwidths of 500 MHz. Data reduction and calibration have been performed using the Data Analysis Application for Radio Continuum (DAARC) code developed by our team (for details, see \citet{2026ApJS..283...69C}). The light curves are then fitted to a quadratic polynomial to remove trends longer than the length of the observation. The sinusoidal curves plotted in Fig.~\ref{fig-lc} are derived by the GLSP method, not by fitting the light curve to a single sine function. Thus, the curve cannot take into account the red noise variability that is not included in any sinusoidal variations.

\section{TESS data} 
TESS \citep{2015JATIS...1a4003R} spans almost the entire sky in the optical waveband with a cadence as rapid as 200 seconds. The monitoring baseline varies from 30 days to $\sim$1 year, depending on the ecliptic latitude of the source. The instrument has four cameras, one above the other, each having a field of view of $24^{\circ} \times 24^{\circ}$. These cameras monitor the sky for one sector, lasting $\sim$ 27 days before moving to a different patch of the sky. In this way, the instrument observed one hemisphere in one year and then inverted to observe the other. As TESS is primarily searching for exoplanets modulating their stars' light, its standard reduction pipeline assumes periodicity while reducing the data, which is not the case for AGNs. In addition, stochastic variability, a salient feature of AGN variability, could be misinterpreted as noise. We have used the Quaver pipeline \citep{2023ApJ...958..188S}, developed to reduce the AGN light curves from TESS. This pipeline not only preserves AGN variability but also removes background flux and systematic uncertainties. 

\section{Archival data} 
The VLBI images and the corresponding flux measurements, spanning the period 1995 to 2026, were obtained from the MOJAVE data archive {\url{https://www.cv.nrao.edu/MOJAVE/}. Since these data are already well calibrated, the images were reconstructed directly using {\tt DIFMAP} \citep{Shepherd_Difmap_1997ASPC..125...77S}. The Swift light curve PKS 0754+100 used in this work covers the period from 2005 to 2026 and is part of the Swift/Burst Alert Telescope (BAT) hard X-ray transient monitoring. It was obtained from the light curve repository, available online at \url{https://swift.gsfc.nasa.gov/results/transients/}.

\section{Significance of QPOs} The periodogram density, $P(f)$, also known as the power spectral density (PSD), of a blazar, usually follows an inverse relation with the frequency $f$, raised to an exponent, $\alpha$, known as the spectral index, which represents stochastic variability \citep{2014MNRAS.445..437M, 2018A&A...620A.185N}. Any QPO, on the other hand, should originate from a possibly deterministic process and should appear above the periodogram density as a peak in Fourier space \citep{2021MNRAS.501.5997T}. Estimating significance with the simulations is very important for identifying false QPO peaks that arise due to the limited observation time or the large amplitude variations at lower frequency owing to the inverse power-law dependence. 

The significance of such quasi-periodic signals is estimated from the p-value assuming the null hypothesis. The null hypothesis is that the variations in the light curves are only caused by the red noise stochastic variations and no QPO signal is present. Here the p-value is the probability of finding a signal stronger than the observed signal and is calculated by estimating the number of peaks with amplitude more than the observed QPO signal. We present our result in terms of significance which is related to p-value as $(1-p)\times 100$\%. 

We have simulated 100,000 light curves with the same statistical (mean, variance) and spectral properties (red noise variation, which goes as $P\propto \nu^{-\alpha}$) as the original observation using the STINGRAY package \citep{1995A&A...300..707T, 2019ApJ...881...39H}. STINGRAY simulates light curves using the number of data points $N$, Total observation time $T$, spacing between the measurement $\Delta T$, and the root mean square of the observation. We earlier binned all of our light curves so that we can use that bin size value to simulate the light curve. So, we achieved the same sampling pattern for both real and simulated light curves at all radio frequencies of 4.8, 15, and 23.6 GHz. To simulate the stochastic variations, we have modeled the power spectrum $P(f)$ of the source with a broken power law, which is basically two simple power laws joined at a break frequency $f_b$, and is expressed as 

\begin{equation}
\begin{split}
    P(f) &= A \left(\frac{f}{f_b}\right)^{-\alpha_{high}} + C,~~~ f > f_b;\\
           &= A \left(\frac{f}{f_b}\right)^{\alpha_{low}} + C,~~~ f < f_b~.
\end{split}
\end{equation} 
Table~\ref{tab:1} lists the best-fit parameters for the broken power law used to model red noise present in radio and optical observations of this source; for the 23.6 GHz data a single power-law produced an adequate fit.

Then, the PSD of each simulated light curve is calculated with a frequency spacing such that the number of frequencies is equal to the number of input points in the given observation \citep{2022ApJ...926L..35O}. 
\begin{table*}
 \centering\hfill
 \caption{Best-fit parameters for the broken power law model for the red-noise present in the observations analyzed in this work. }
\begin{tabular}{|c|c|c|c|c|}
\hline
Observation &$\alpha_{low}$ & $\alpha_{high}$  & $f_b$ (d$^{-1}$)  & A \\\hline
4.8 GHz & 0.33 $\pm$ 0.01 &  1.89 $\pm$ 0.22 & 0.0035 $\pm$ 0.0002 &0.10 $\pm$ 0.01\\ \hline
23.6 GHz & 0.71 $\pm$ 0.06 &  0.71 $\pm$ 0.06 & - & 0.11 $\pm$ 0.01\\ \hline
TESS & 0.32 $\pm$ 0.02 &  1.51 $\pm$ 0.1 & 0.55 $\pm$ 0.03 & 0.02 $\pm$ 0.01\\ \hline

\end{tabular}\label{tab:1}

\end{table*}
The frequency range considered in this work is 0.001--0.01 days$^{-1}$ for radio data, since that is the range where the convincing QPO signal can be found, given that the temporal baseline of the observation is $\sim$3000 days. The PSD at higher frequencies is greatly affected by Poisson noise and those at lower frequencies can be  misinterpreted as QPO signals due to the inverse P-$\nu$ relation.
The local significance of such a peak can be estimated by the number of periodogram densities below the claimed periodic-peak at a particular frequency. If the number of periodogram densities is estimated for all frequencies considered for analysis, then the significance becomes global \citep{2026MNRAS.545f2211T} for a single object. For this global significance, the frequency grid is chosen such that the spacing between the frequencies is 1/($\zeta$T), where $T$ is the total duration of the observation. The value of $\zeta$ varies between 1 and 5 such that the number of frequencies is equal to the number of data points \citep{1982ApJ...263..835S} and is further used to construct a frequency grid between frequencies $1/T$ and $N/2T$, where $N$ is the number of data points in the observation. We then search simulations estimated for these frequencies which exceed the observed peak in the frequency range 0.001--0.01 d$^{-1}$, where the red noise variations are valid. 

Since this source is selected from 11 variable sources in the sample, the number of sources searched for possible periodicity, in addition to the number of frequencies examined, should also be taken into account to calculate the full global significance \citep{2025MNRAS.541.2955P}. Including these other sources takes into account this aspect of the ``look elsewhere effect''. The full global p-value $p_{g}$ can be expressed in terms of local significance  $p_{l}$ as \citep{2026MNRAS.545f2211T}
\begin{equation}
    p_{g} = 1 - (1 - p_{l})^T,
\end{equation}
where now $T$ is the trial factor, defined as  
\begin{equation}
    T = M \times S.
\end{equation}
Here $M$ and $S$ respectively are the number of independent frequencies and the number of blazars searched for possible periodicity. For this work, the value of $S$ is 11. The number of independent frequencies is calculated using the grid spacing defined above for the frequency range of 0.001-0.01 d$^{-1}$ \citet{2022ApJ...926L..35O}. Thus, the value of $M$ is estimated to be 30 for a temporal baseline of 3000 days. Putting in these values, the local significances of 4.51$\sigma$ and 4.89$\sigma$, which are calculated for {the two radio frequencies}, correspond to global significances of 2.98$\sigma$ and 3.50$\sigma$ respectively. 

For TESS data, the significance is calculated in the frequency range of 0.1--1 d$^{-1}$, where the red noise variations are not affected by lower- and higher-frequency complications as described above. The second part of the look elsewhere effect is not applicable to the TESS observations used in this work as it was examined without searching for possible periodicity in any other objects, but rather was checked only after periodicity was found in the radio observations of this source.

\section{Correlation studies} The auto-correlation function (ACF) is the discrete correlation function (DCF) of a light curve with itself, and its first significant peak may indicate the presence of a quasi-periodic signal in the observation \citep{1988ApJ...333..646E,jankov_isidora_2022_7253034}. For a particular radio band, we simulate 100,000 light curves with the similar properties of real observation and then calculate the ACF for all these light curves. So, for each time lag, we have the distribution of ACFs. Then, we estimate the ACF value which is greater than 99.99\% (4$\sigma$) of all ACF values at a particular time lag. When calculated for all time lags, this gives us the 4$\sigma$ significance level for the ACF of a given light curve. The ACFs of the 4.8 and 23.6 GHz light curves show positive values, with more than  4$\sigma$ significance, at respective time delays of 269$_{-9}^{+7}$ and 271$_{-3}^{+3}$ days, in agreement with the results obtained with GLSP and WWZ. We also calculated the cross-correlation function (CCF) between 4.8 and 23.6 GHz for segment 1, where the  23.6 GHz observations lead the 4.8 GHz observations by  22$_{-4}^{+6}$ days (see Fig.~\ref{fig-ccf}).

\begin{figure*}[t]
    \centering
    \includegraphics[width=1.0\linewidth]{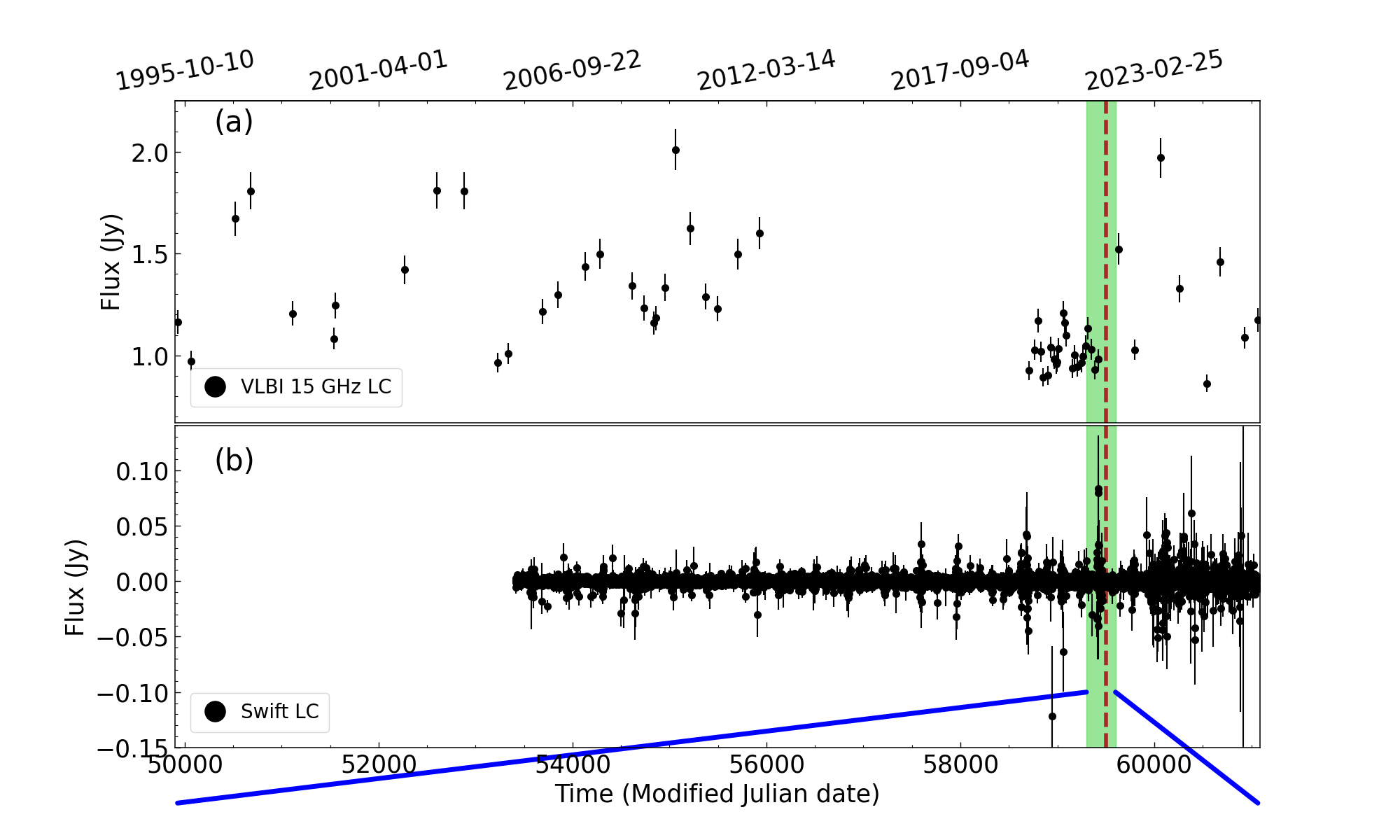}
    \includegraphics[width=1.0\linewidth]{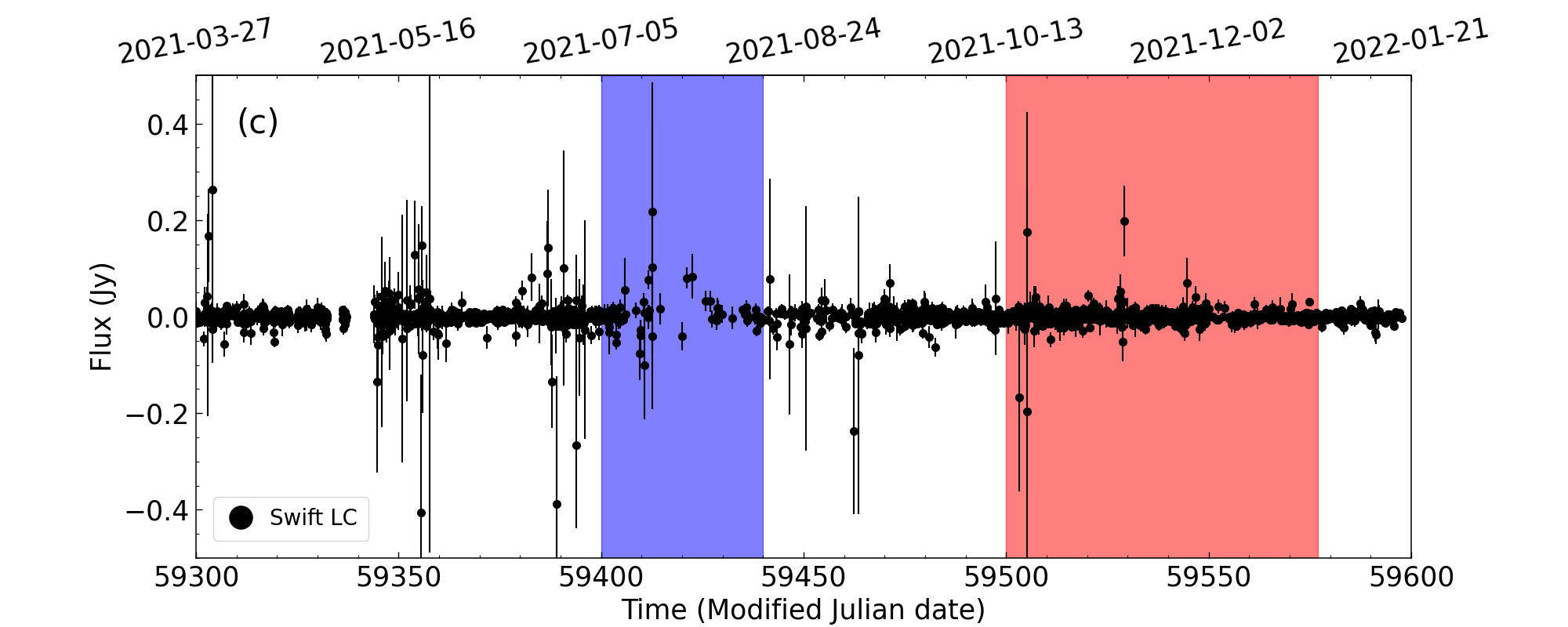}
    \caption{Complete available flux measurements from VLBA and Swift observations of PKS 0754+100 in panels (a) and (b), respectively. Up to  2012, the 15 GHz observations were in a high-variability state with no strong variations present in the X-ray light curve. The highest flux measurement for Swift was observed around MJD 59400, at the onset of another high-variability state. After MJD 59500, strong fluctuations in flux can be seen in the Swift light curve along with a high- variability state at 15 GHz (and also at the 23.6 GHz SMMAN observations). Panel (c) shows the zoomed-in Swift light curve for MJD 59300--59600. This interval marks the change in low to high variability state in radio observations, which is also shaded in green in panels (a) and (b). The blue region highlights the region with the flare, which lasted for about 40 days. This could be related to the QPO observed in optical TESS observations for the time shaded in crimson red and subsequently to the periodic modulations in radio wavebands. }
    \label{fig-swift}
\end{figure*}
\begin{figure*}
    \hspace{2cm}\includegraphics[width=0.8\linewidth]{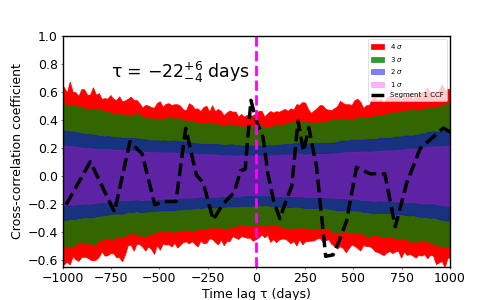}
    \includegraphics[width=0.5\linewidth]{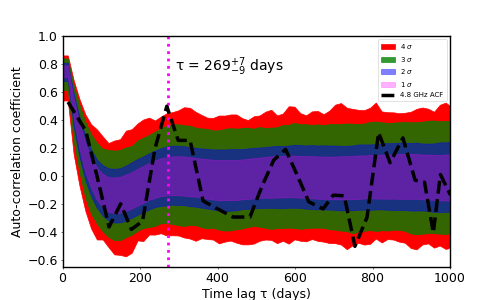} \includegraphics[width=0.5\linewidth]{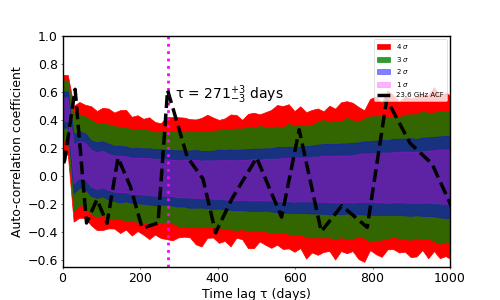}

    \caption{Cross-correlation studies for the 4.8 and 23.6 GHz light curves. The black dashed curves in each plot show the correlation function. The magenta, green, blue, and red regions correspond to 1$\sigma$, 2$\sigma$, 3$\sigma$, and 4$\sigma$, respectively. The upper panel shows the cross-correlation function (CCF) between 4.8 and 23.6 GHz observations for the segment before MJD 59500. The time lag is negative, indicating that the emission at 23.6 GHz leads that at 4.8 GHz. The left plot of the bottom panel shows the auto-correlation function (ACF) for 4.8 GHz, and a 4$\sigma$ significant correlation is found at the time delay of 269 days, which is in agreement with the period found with the GLSP and WWZ methods. The right plot shows the ACF for the first segment of 23.6 GHz, where a significant correlation is found at 270 days, also in agreement with the results obtained from other methods.}
\label{fig-ccf}
\end{figure*}

    \begin{figure*}
    \includegraphics[width=1.0\linewidth]{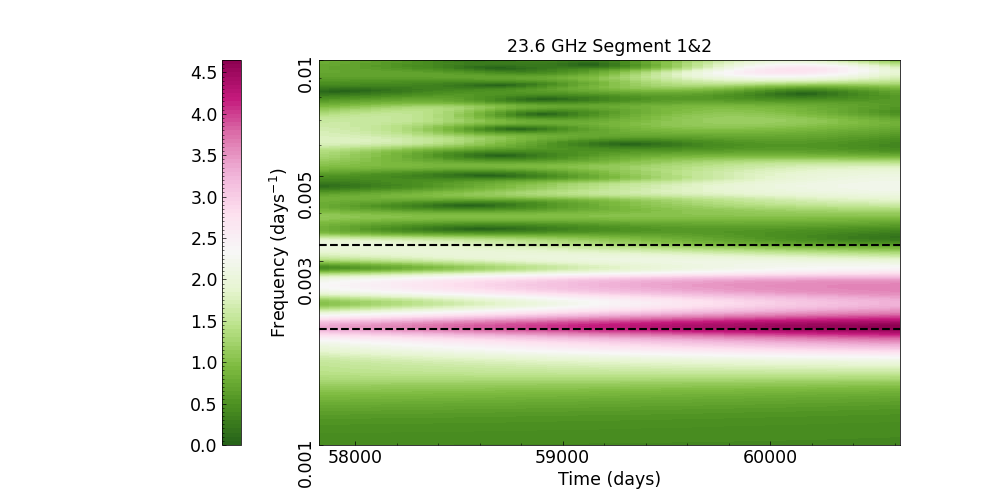}
    \caption{The WWZ colour-colour diagram for 4.8 GHz between MJD 57700 and 60620. The square root of WWZ power is plotted to better visualize weak signals.
    A strong concentration of WWZ power is observed at the frequency of 0.002 d$^{-1}$, corresponding to the period of 494 days. A very weak signal is found to be around 285 days which changes the frequency as the observation advances and gain strength at the end but still very weak as compared to the signal at 494 days. So, the change in variability state abruptly affects the QPO signal in this segment and introduces another strong  QPO signal around 490 days.     }
   \label{fig:k}
\end{figure*}






\bibliographystyle{aasjournalv7}
\bibliography{chapter}

\end{document}